\documentclass[10pt]{article}

\usepackage[dvips]{graphicx}  
\usepackage{amsmath}
\usepackage{amssymb}
\usepackage{amsbsy}
\usepackage{amsfonts}
\usepackage{latexsym}
\renewcommand{\O}[0]{\mathrm{O}}

\newcommand{\im}[0]{\mathrm{Im}}

\newcommand{\const}[0]{\mathrm{const}}
\renewcommand{\leq}[0]{\leqslant}
\renewcommand{\geq}[0]{\geqslant}

\renewcommand{\hat}{\widehat}
\begin{document}

\title{Scattering, trapped modes and guided waves in waveguides and diffraction gratings}
\author{V.E.Grikurov\\ Department of Mathematical and Computational Physics,\\
 St.Petersburg University, Russia\\
e-mail: grikurov@math.nw.ru}
\date{}
\maketitle
\thispagestyle{empty}

\section{Scattering matrices and existence of trapped waves}

\begin{figure}[ht]
\includegraphics[width=.2\textwidth]{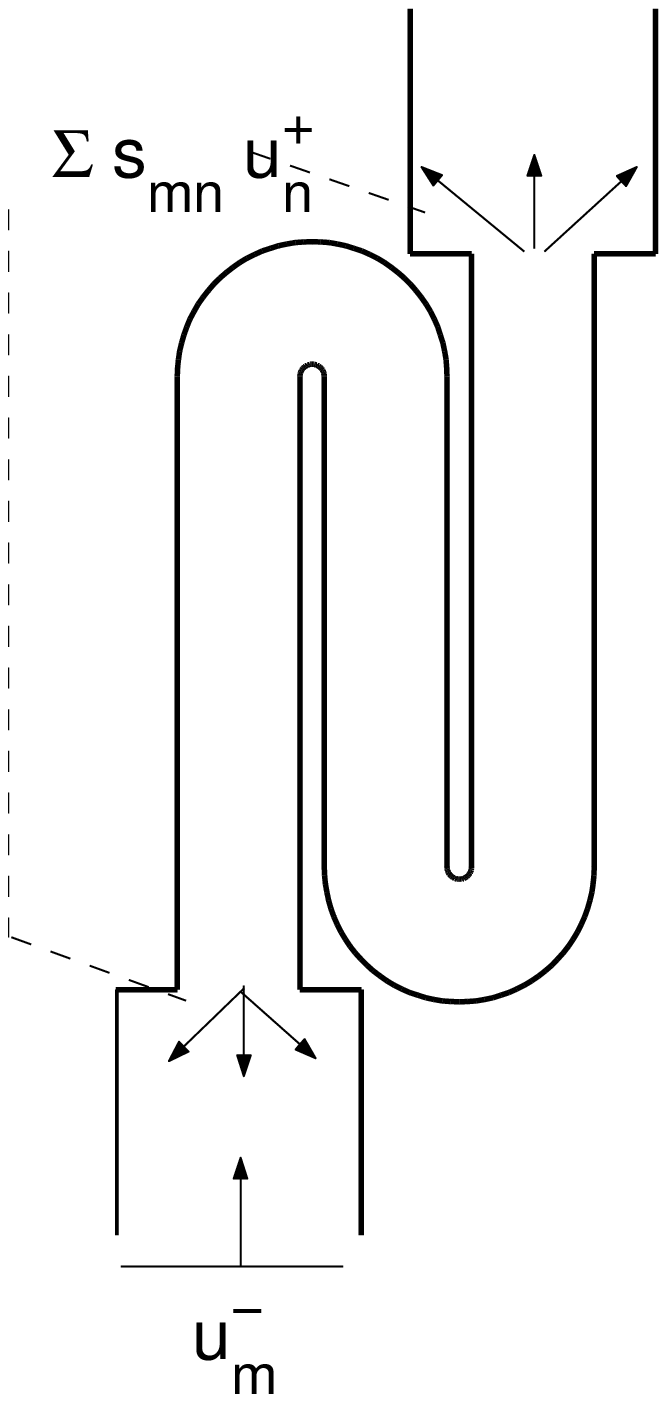}
\hspace*{0.25cm}
\includegraphics[width=.4\textwidth]{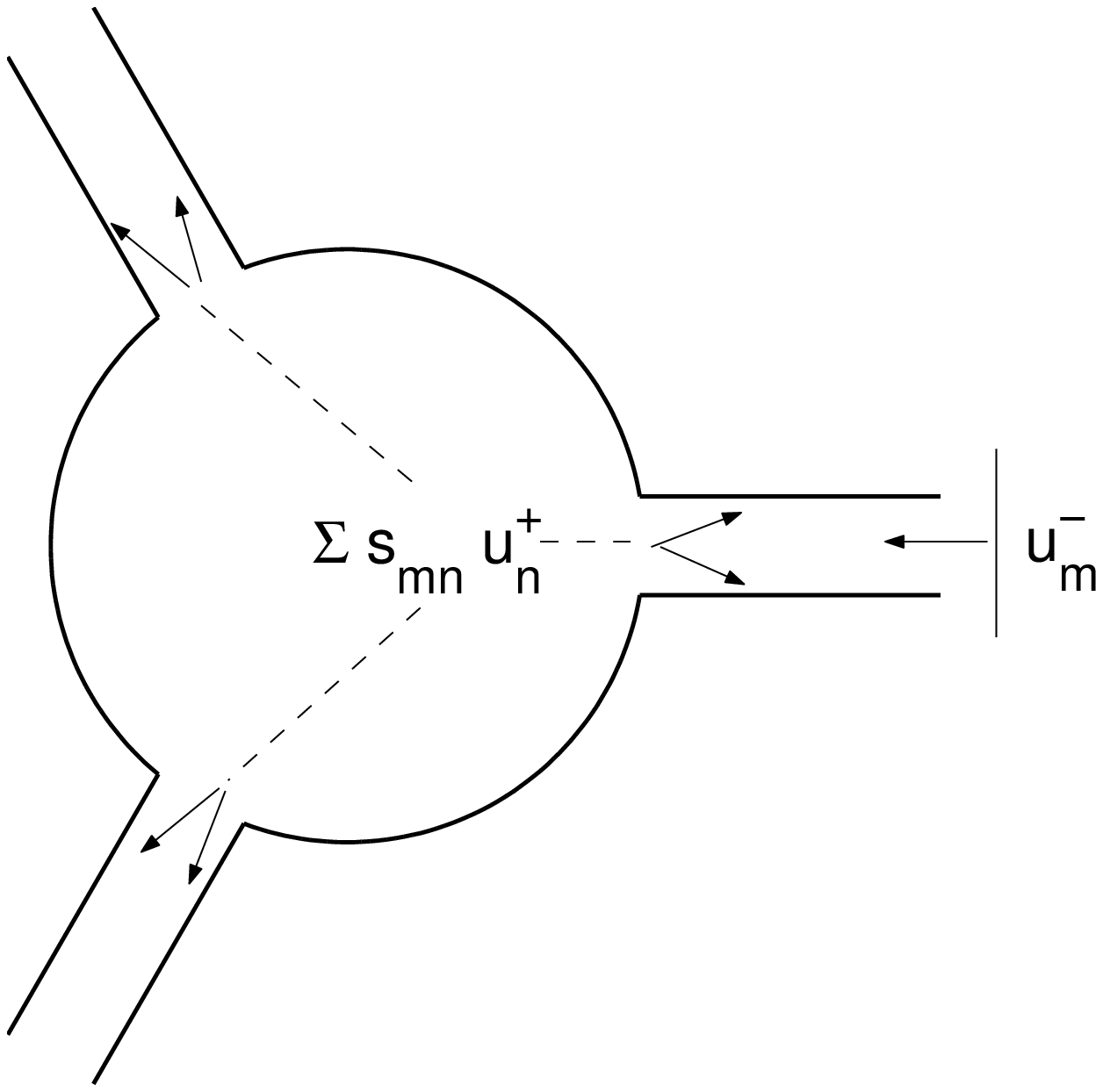}
\includegraphics[width=.35\textwidth]{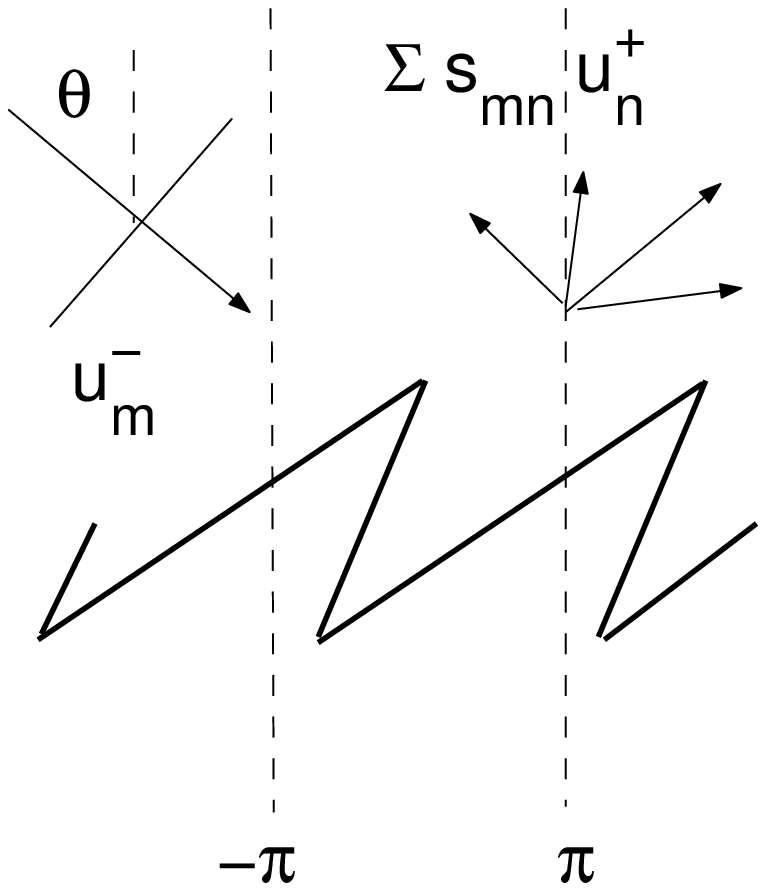}
\caption{\small Various systems with nontrivial scattering properties:
``quantum wire'', open resonator, diffraction grating. Dialing with
the grating, one can reduce the problem to the strip whose width is equal to the grating's
period ($2\pi$ in the figure) with the quasi-periodical boundary conditions
(\ref{qpc}).
 All these systems can support trapped
modes or guided waves.}
\label{ex}
\end{figure}

Systems with finitely many scattering channels are discussed in various branches of
wave physics. One can mention, for instance, diffraction gratings (optics), waveguides
with local perturbations (acoustics, microwave and quantum physics), etc.
(few examples  are shown in Fig.\ref{ex}). In all cases,
far away from the scattering area both incident and scattered fields are sums
over ``modes'' (specification of modes is given below).
By another words, assuming some throughout
enumeration of modes over all channels, any solution to the scattering
problems asymptotically behaves as
\begin{equation}\label{A}
\psi\underset{|x|\to\infty}{\sim}\sum_n\left(c^-_n u_n^-+ c^+_nu_n^+\right)
\end{equation}
with some coefficients $c^\pm_n$.
\vskip 1cm

In what follows let us restrict ourselves
either by the Helmholtz $\Delta\psi+k^2n^2(x)\psi=0$
 or by the Schr\"{o}dinger $-\Delta\psi+V(x)\psi=E\psi$ equation and assume that
for $|x|\geq R_0$ $n(x)\equiv 1$ or  $V(x)\equiv 0$, respectively. Then
modes $u_n^\pm(x)$ are solutions to the homogeneous problem for $|x|\geq
R_0$. For example, the waveguide modes are
 $u_n^\pm=\phi_n(x_\bot)\,e^{\pm i\lambda_n z}/\sqrt{\lambda_n}$,
$\lambda_n=\sqrt{k^2-\mu_n}$ (or with $E$ instead of $k^2)$, where $x=(z,x_\bot)$, $z,x_\bot$
are local  coordinates respectively along (outward to infinity) and across each
channel, $\{\mu_n,~\phi_n(x_\bot)\}$ are eigenpairs to the problem in channel's
cross-section $z=\const$;
the gratings modes are
$u_n^\pm=\exp\{\pm i\lambda_n z + (n+\alpha)y\}/\sqrt{4\pi\lambda_n}$,
$\lambda_n=\sqrt{k^2-\left(n+\alpha\right)^2}$ ($x_\bot=y$ is one-dimensional here)
satisfy the quasi-periodical boundary conditions
\begin{equation}\label{qpc}
\left.u\right|_{y=\pi}=e^{2\pi i\alpha}\left.u\right|_{y=-\pi}\,,\quad
\partial_y\left.u\right|_{y=\pi}=e^{2\pi i\alpha}\partial_y\left.u\right|_{y=-\pi}
\end{equation}
(here grating's period equals to $2\pi$ and  $\alpha=-k\sin\theta$, $\theta$ is the angle
of incidence, see Fig. \ref{ex}). It is also assumed in (\ref{A}) that each mode
differ from zero only at ``its'' channel: by $u^\pm_n$ we
actually denote
\begin{equation}\label{cof}
u^\pm_n\leadsto u^\pm_n(z,x_\bot)\eta(z)\,,
\end{equation}
where $\eta(z)$ is a smooth cut-off function which equals to one inside the
channel and zero elsewhere.

Modes with numbers $n$ such that $\lambda_n>0$ propagate (oscillate) along the channel
at which they are defined; we distinguish \textit{outgoing} and
\textit{incoming} (respectively, $u_n^+$ and $u_n^-$ in our notations) modes
by the sign of the energy flux density
$\im\left(\frac{\partial u}{\partial \nu}\overset{-}{u}\right)$, where $\nu$ is the
outward normal to channel's cross-sections. At the threshold frequencies
$\lambda_n=0$, and  the above definition of modes failed; following the paper
\cite{NP} we define threshold (standing) modes by the following replacement:
$e^{\pm i\lambda_n z}/\sqrt{\lambda_n}\leadsto\left(1\mp z\right)/\sqrt{2}$.

The sum in (\ref{A}) is normally taken over all propagating modes
(including standing ones). Let the total number of such modes is $2N$.
Then the space of solutions with asymptotics (\ref{A}) is $N$-dimensional
(see \cite{NP} again) and, consequently, there exists $N$ linearly
independent rows $\left(c_1^-,\dots,c_N^-,c_1^+,\dots,c_N^+\right)$, which
form two $N\times N$ matrices $\mathbb{C}^\pm:=\|c_{mn}^\pm\|$.
Both $\mathbb{C}^\pm$ are invertible (otherwise nontrivial solution to homogeneous
problem with an asymptotics containing only outgoing or only incoming modes
can exist). The unitary matrix $\mathbb{S}:=\left(\mathbb{C}^-\right)^{-1}\mathbb{C}^+$
is called ``scattering matrix''. The question of its numerical determination
was discussed in a long series of publications (see, e.g., \cite{G,ES} as a
most general method).
The approach suggested below can be, in particular, applied to that end as
well (see Section \ref{3}).
However, our goal is a little bit wider.

Let us agree that $\im\lambda_n\leq 0$ and include into the sum (\ref{A})
all modes corresponding to $|\im\lambda_n|<\gamma$ for some positive $\gamma$.
\footnotemark[1]\footnotetext[1]{It is known \cite{NP} correction term
  is  $\O\left(e^{-\gamma|x|}\right)$ this case.}
Note that under our choice of the sign of $\im\lambda_n$
modes $u_n^-$ are decay whilst $u_n^+$ are grow as $|x|\to\infty$, $n>N$.
Now the question of interest is: \textit{\textbf{is there exist linear combination
of solutions (\ref{A}) containing only decaying exponents?}} Existence of such a
combination would indicate the solution to the homogeneous problem
either localized at compact domain (trapped modes or bound states to waveguides
and resonators) or propagating along a grating groove and decaying away from the grating
 (guided or surface grating waves).

Suppose the new matrix $\mathbb{C}^+$  is
at our disposal and write it down as
$$\mathbb{C}^+=\left(\begin{array}{cc}
  C_{(11)} & C_{(12)} \\
  C_{(21)} & C_{(22)}
\end{array}\right)\,,$$
where the block $C_{(11)}$ is of size $N\times N$ and corresponds to
the coefficients are in front of propagating modes.
 By straightforward verification one concludes that the condition
\begin{equation}\label{C}
  \det C_{(22)}= 0
\end{equation}
is sufficient for the existence of the aforementioned decaying solution.
Consider (\ref{C}) as the equation for the wave frequency $k$ (or particle energy
$E$ in quantum physics). Thus we arrive to the
 existence criterion for localized solutions.

 It can be also shown that scattering matrix $\mathbb{S}$ is now the left-upper
 block of the product  $\left(\mathbb{C}^-\right)^{-1}\mathbb{C}^+$. Thus
 matrices $\mathbb{C}^\pm$ contain information about both scattering and
 trapping properties of the system.

However, the numerical determination of the matrices  $\mathbb{C}^\pm$ is not
a trivial task since the asymptotics
(\ref{A}) contains
exponents that grow and decay with different rates. The leading term dominates, and
all coefficients cannot be found accurately. The
key point of the  approach suggested in the next Section is to avoid this difficulty.
In the Section \ref{4} we give a very brief review on previously known
numerical approaches and provide new examples of trapped modes.

\clearpage
\section{Description of the approach}\label{2}

 The main computational idea is as follows.

Consider the truncation of an infinite domain  by removing infinite parts
of all channels, starting from some distance $R>R_0$ (i.e., truncating lines
are $z_j=R$, $j$ is the channel's number). Let $\hat{\psi}$ be the solution to the
auxiliary problem which inherits all conditions of the original problem (at an
infinite domain) with the following artificial condition at
$z_j=R$:
\begin{equation}\label{abc}
  \mathcal{B}_j\left(\hat{\psi}-\sum_n\left(\hat{c}^-_n u_n^-+ \hat{c}^+_nu_n^+\right)
  \right)=0\quad\forall j\,,\quad \mathcal{B}_j:=
  \left.\frac{\partial}{\partial z_j}+i\zeta\right|_{z_j=R}\,,
\end{equation}
where $\zeta>0$ and coefficients $\hat{c}_n^\pm$ are arbitrary.

Conditions (\ref{abc}) came into scene as the differentiation of
 the asymptotics (\ref{A})  (positive $\zeta$ is necessary to avoid possible
coincidence of $k^2$ with
the spectrum of the auxiliary problem). Let us now try to chose coefficients $\hat{c}^\pm_n$ in such a way that
$\hat{\psi}$ approximately agrees with the right-hand part in (\ref{A}) at
$z_j=R$. Thus one can expect that for sufficiently large $R$ the numbers
$\hat{c}^\pm_n$  are good approximations to $c_n^\pm$.

By this motivation we arrive to the condition on unknowns $\hat{c}^\pm_n$:
choose these coefficients in such a way that
\begin{equation}\label{F}
\sum\limits_{j}\int\limits_{z_j=R}
\left|\hat{\psi}-\sum_n\left(\hat{c}^-_n u_n^-+ \hat{c}^+_nu_n^+\right)
\right|^2 dx_\bot\mapsto\min\,.
\end{equation}
It can be shown that the left-hand part in (\ref{F}) goes to zero
as $R\to\infty$ with exponential rate and
\begin{equation}\nonumber
\sum_{n}\left|\,\hat{c}^\pm_n-c^\pm_n\right|^2\underset{R\to\infty}{=}
\O\left(\,e^{-\hat{\gamma} R}\right)\,.
\end{equation}
The proof of the similar convergence for the entries of the scattering matrix $\mathbb{S}$,
 as well as the estimation of the constant $\hat{\gamma}$, can be found
in \cite{GHNP1,KNP1}.

From (\ref{F}), one has the simple algorithm to obtain approximations
$\hat{c}^\pm_n$ to $c^\pm_n$ that are sought for. In fact, the functional (\ref{F})
is quadratic with respect to $\hat{c}^\pm_n$. The Hermitian matrix of its coefficients
can be written as
\begin{equation}\label{matr}
\left(
  \begin{array}{cc}
    \hat{A}^- & \hat{B} \\
    \hat{B}^* & \hat{A}^+
  \end{array}
\right)\,,
\end{equation}
where
$$\hat{A}^\pm_{pq}=\sum\limits_{j}\int\limits_{z_j=R}
\left.\overline{\left(\hat{u}_p^\pm-u_p^\pm\right)}
\left(\hat{u}_q^\pm-u_q^\pm\right)\right|_{z_j=R}dx_\bot\,,$$
$$\hat{B}_{pq}=\sum\limits_{j}\int\limits_{z_j=R}
\left.\overline{\left(\hat{u}_p^--u_p^-\right)}
\left(\hat{u}_q^+-u_q^+\right)\right|_{z_j=R}dx_\bot\,,$$
and the functions $\hat{u}_n^\pm$ are the solutions to the problems with artificial
conditions $\mathcal{B}_j\left(\hat{u}_n^\pm-u_n^\pm\right)=0$.

Note that the
data to the auxiliary problems on the differences $\left(\hat{u}_n^\pm-u^\pm_n\right)$
don't contain growing as $R\to\infty$
exponents as well as summation of different rate exponents.
These problems have to be solved numerically by means of any
appropriate method (e.g., finite element method).

Finally, given the matrix (\ref{matr}) one can find its spectrum and select
$M$ smallest magnitude eigenvalues and corresponding eigenvectors
$h_1,\dots,h_M$ (here $2M$ denotes the size of the matrix) and put
\begin{equation}\label{R}
  \left(
  \begin{array}{c}
    \hat{\mathbb{C}}^- \\
    \hat{\mathbb{C}}^+
  \end{array}\right) =   \left(h_1,\dots,h_M\right)\,.
\end{equation}
This finalize the description of the computation procedure. It is clear that
the approach doesn't sensitive to the geometry of a problem.

\clearpage
\section{Examples of scattering data computation}\label{3}

In this section we discuss the the application of the above approach to the
computation of scattering data.

\subsection{Quantum control on electron stream}

\begin{figure}[ht]
\includegraphics[width=.35\textwidth]{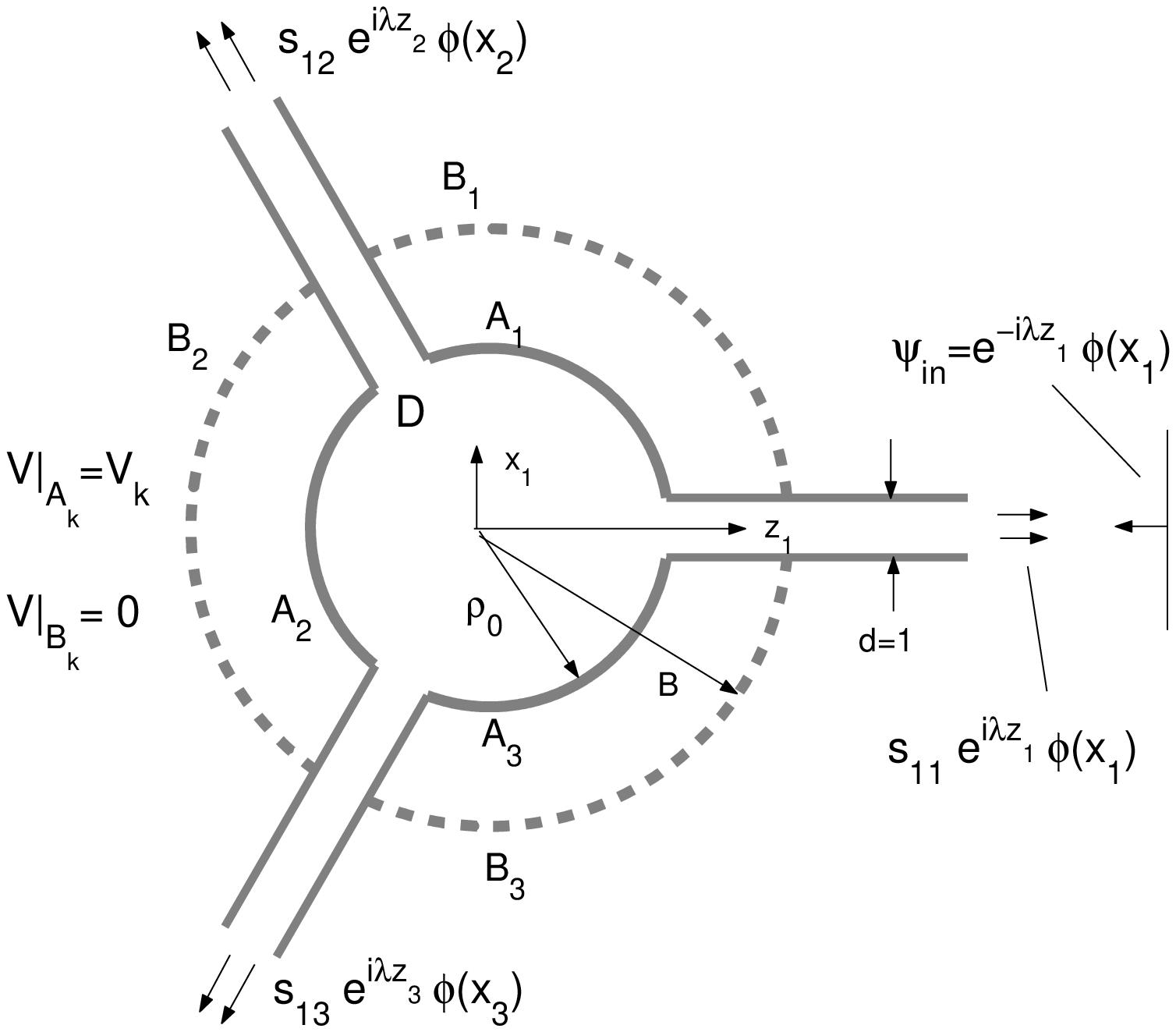}
\hspace*{0.5cm}
\includegraphics[width=.5\textwidth]{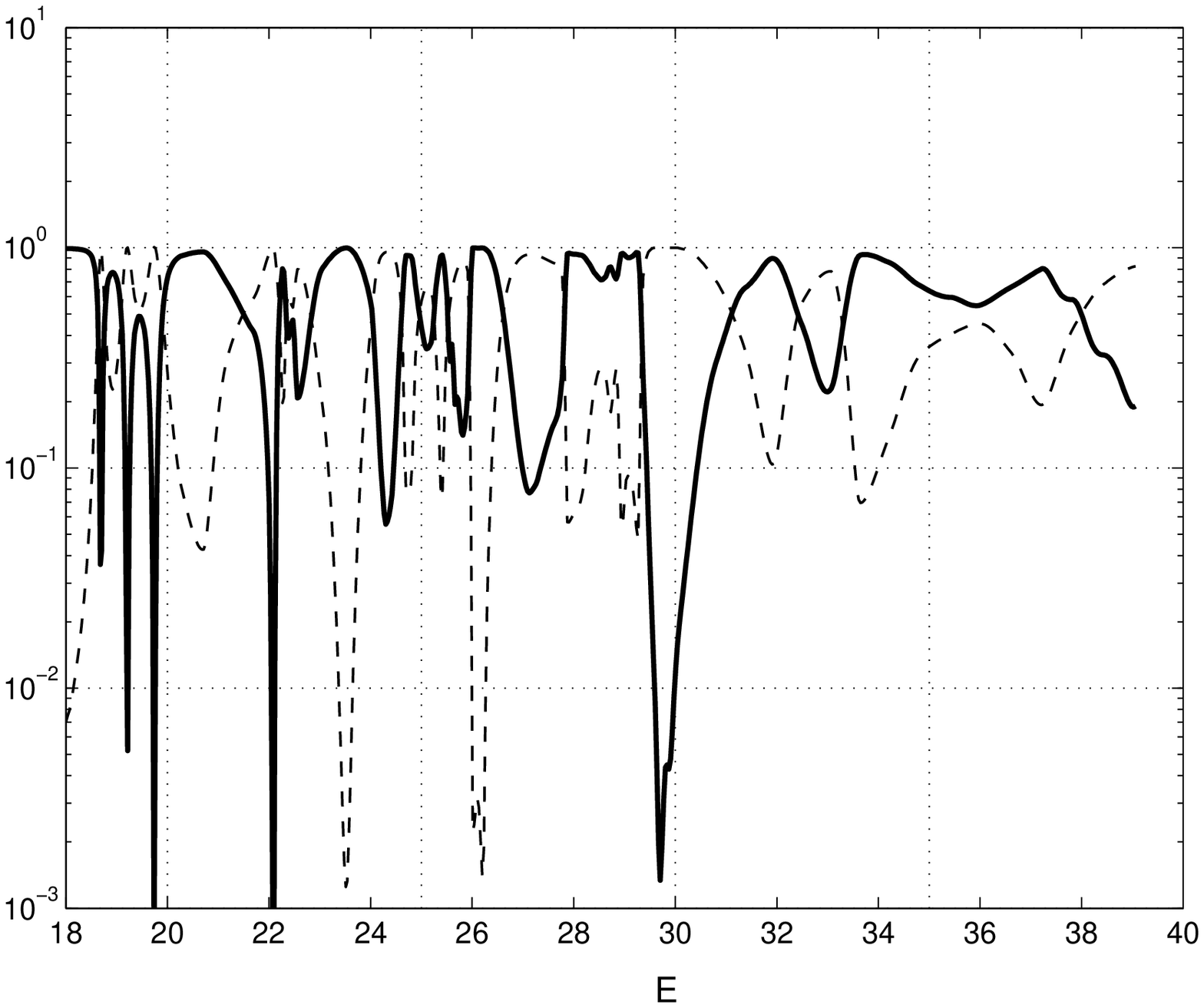}
\caption{\small  \textit{Left:} Sketch of the model.
\textit{Right:}
Transporting losses $|s_{11}|^2+|s_{13}|^2$ (solid line) and the transmission coefficient
$|s_{12}|^2$ (dashed line) versus energy $E$ of an incident electron; handling
potentials are fixed as $V_1=0$, $V_2=V_3=(1.5\pi)^2\sim 22.2$; the size of
resonator $\rho_0=3d$.}
\label{Trigger}
\end{figure}

Consider the two-dimensional domain $D\subset\mathbb{R}^2$  consisting of a resonator
(i.e., of a disk of radius  $\rho_0$) which is connected to infinity by means
of three straight channels (waveguides) of width $d$ (see the left part in Fig.\ref{Trigger});
directions of axes
of two channels (say, 2nd and 3rd) are  symmetric with respect
to the direction of the 1st channel axis (the symmetry is not essential and is introduced
to decrease the number of problem parameters).
We assume that the motion of an electron is bounded by the domain $D$ and
the wave function $\psi(x,y)$ vanishes at the boundary of $D$.
\footnotemark[2]
\footnotetext[2]{The Schr\"{o}dinger equation is written in reduced units
when unit of length is $d$
and the unit of energy is $\hbar^2/(2m^*d^2)$,  $m^*$ is an effective electron mass.}

 The governing potential $V$ is induced in the following way.
 \footnotemark[3]\footnotetext[3]{This model was suggested by Prof. L.M.Baskin
 (St.Petersburg, Russia);
 more detailed description of the model and extended results are in press \cite{BGNP}.}
 Let the resonator's walls $A_1$, $A_2$, $A_3$ are charged by potentials
$V_1$, $V_2$, $V_3$, respectively; the handling is realized by variation of
values $V_{1,2,3}$. The whole system is shielded by three non-closed lines
$B_1$, $B_2$, $B_3$, each shield $B_j$ consists of a segment of radius $B$ and two
rays directed along channel's walls (these shields are shown in
Fig.\ref{Trigger} by dashed lines). Thus the potential $V$
is the solution to Laplace equation in the unbounded domain $D_B$ (restricted
by the shields $B_{1,2,3}$) and satisfies the Dirichlet boundary conditions
$\left.V\right|_{A_j}=V_j$, $\left.V\right|_{B_j}=0$. It is known
that such solution exponentially decays along the channels. It allows to
accept, as an approximation to $V$, the solution to the problem in a
finite part of $D_B$ that is located inside the disk of sufficiently large
radius $B$ (with the zero boundary condition at new boundaries).

We consider the scattering problem in the domain $D$ under the energy range
$\pi^2<E<(2\pi)^2$ (in reduced units), that is, between first and second
thresholds and, consequently, the size of the scattering matrix is $3\times 3$
(one propagating mode at each channel). The examined  energy range corresponds to
 realistic energy values $(0.01\div1)\,$ev for $d=(1\div10)\,$nm.

Let the incident electron stream comes along the 1st channel, so the
scattering data is collected in the first row $\|s_{1j}\|$, $j=1,2,3$, of
the scattering matrix $\mathbb{S}$.
Now the question is if one can choose any combination of the
energy $E$ and handling potentials  $V_1$, $V_2$, $V_3$ such that the
transport of an electron stream  either to 2nd or to 3rd channel is close to
certain
(i.e., either $\left|s_{12}\right|^2\approx 1$ or $\left|s_{13}\right|^2\approx 1$)?
If yes, the transport can be switch from second to
third channel by trading places of $V_1$ and $V_3$.

Some of the results are given in the right-part in Fig.\ref{Trigger}. It is seen that
transporting losses can be reduced below 0.1\%
 \footnotemark[4]\footnotetext[4]{The accuracy of computation of the
 scattering coefficients
better than $0.005$; thus, the total transporting losses, i.e.,
$|s_{11}|^2+|s_{13}|^2$, exceeding $(2\div5)10^{-5}$ were found safely.}
by variation of $E$ (the same effect can be achieved by variation of $V$ as well,
not shown here).

\subsection{Conductance of bent waveguides}\label{3.2}

\begin{figure}[ht]
\hspace*{1cm}
\includegraphics[width=.25\textwidth]{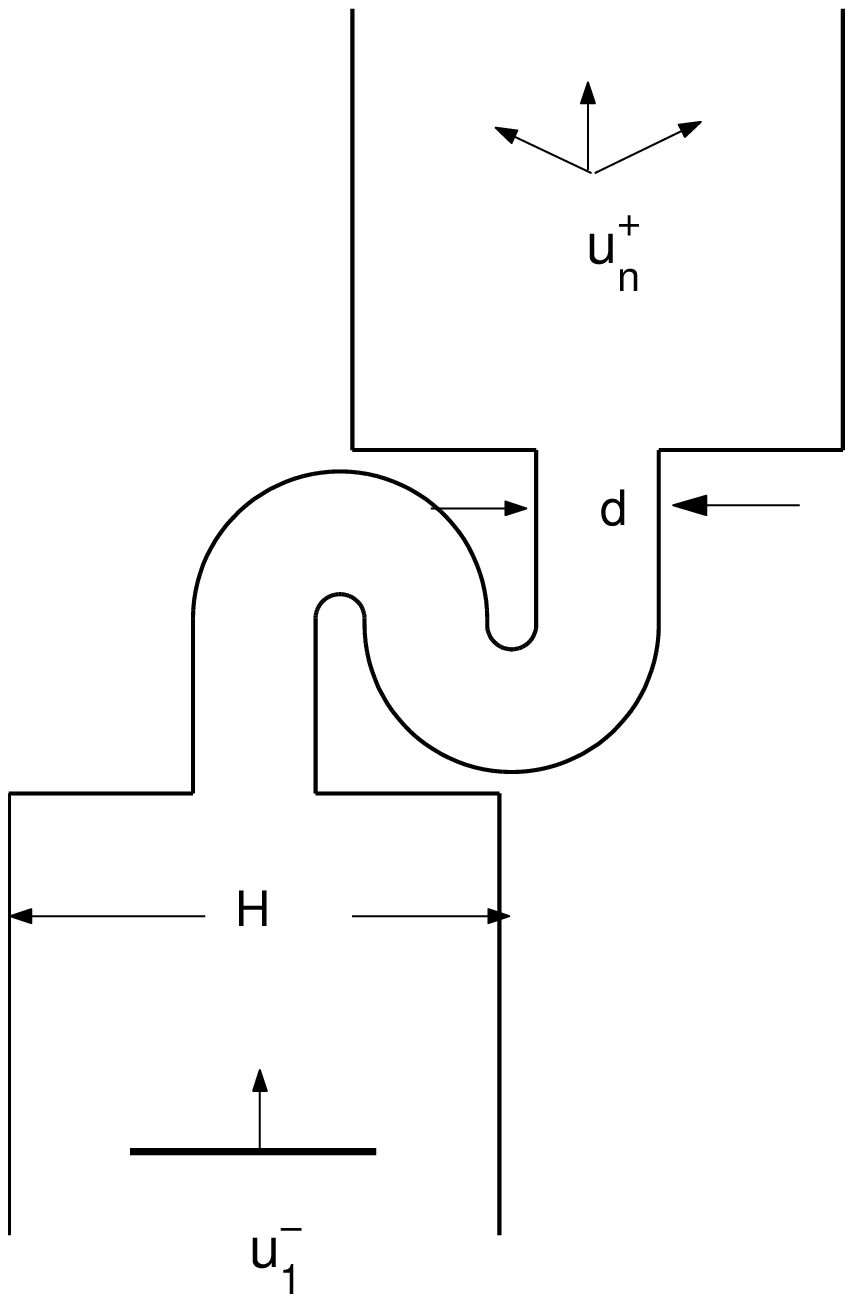}
\hspace*{2cm}
\includegraphics[width=.5\textwidth]{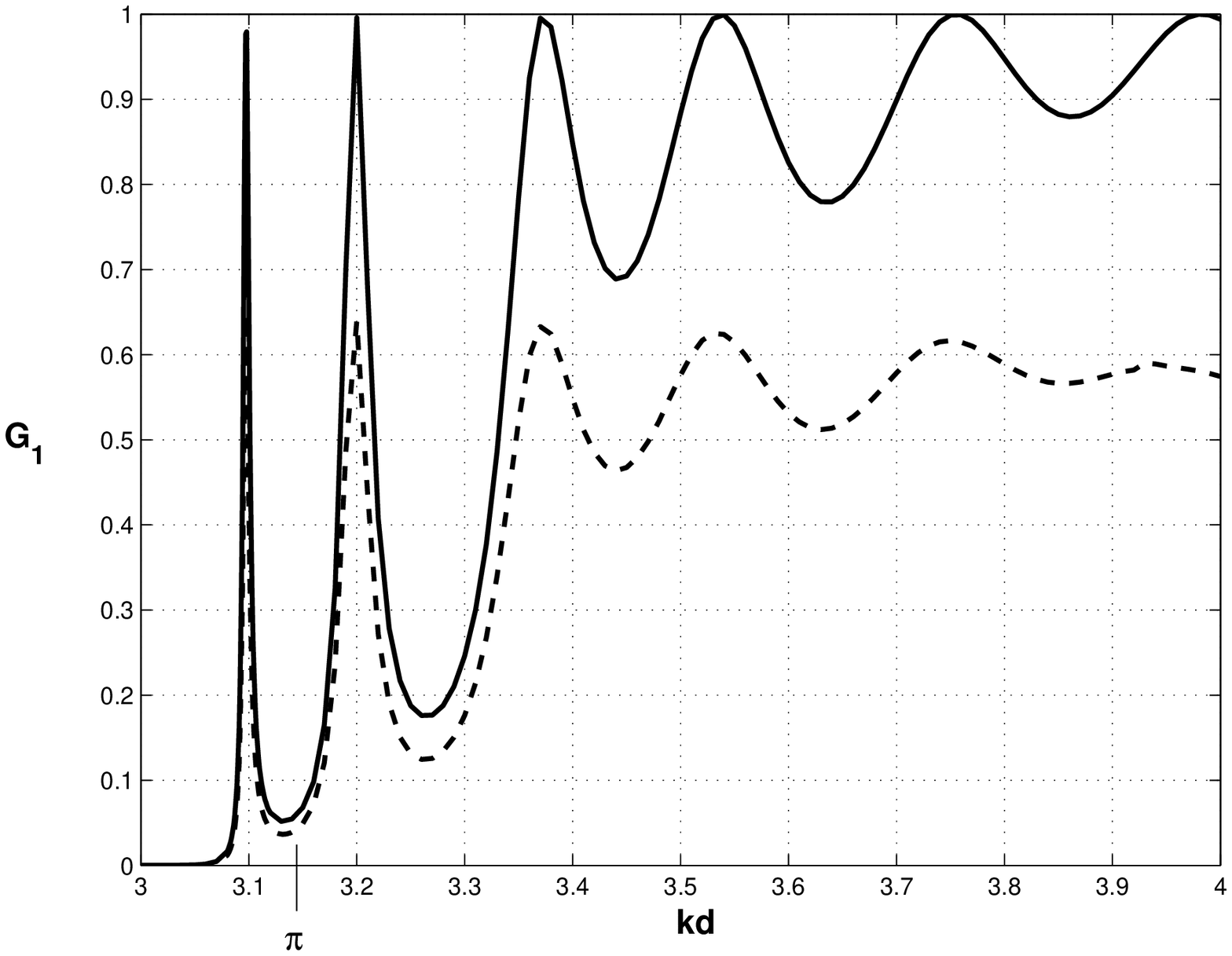}
\caption{\small  \textit{Left:} A bent waveguide (numerous dimensions
of the bend are not specified) of width $d$
connected with two open leads of width $H$;
\textit{Right:}
Normalized conductance $G_1$ (due to the lowest mode in the lead)
versus dimensionless frequency $kd$: solid line --- $H/d=2$, dashed line ---
$H/d=4$. Note that the location of conductance maxima doesn't depend on the ratio $H/d$ and
the first peak is observed (below the bent waveguide threshold $\pi$)
at $kd\approx 3.097$ (compare with the trapped mode frequency
for the  waveguide with the same bent part, see Fig.\ref{Trapped}).}
\label{Snake}
\end{figure}

Two-dimensional bent waveguides are often considered as models of microwave
devices or quantum wires (see \cite{BGRS}). Scattering problem for such a model
is to determine the matrix $\mathbb{S}$
corresponding to the solution of the free Schr\"{o}dinger (Helmholtz)
equation in a curved strip with asymptotics (\ref{A}) and
satisfying Dirichlet boundary conditions. Normally the implementation of
a device presumes the strip of piecewise constant width (e.g., a bent
waveguide is connected to open leads, as shown in the left part of
Fig.\ref{Snake}); this case the definition of modes and scattering matrix
is related to leads.

 Typically the wave process in a system is energized by an incoming single
  mode (number $m$) of inlet lead, and the observed quantity is the \textit{normalized conductance}
 $G_m=\sum_{n}\left|s_{mn}\right|^2$, where sum is taken over all outgoing
 modes of outlet lead.

It is clear that modes can propagate throughout the leads at frequencies below
the threshold ($kd=\pi$) of the bent strip. Transmission of energy at such
frequencies through the bend is classically forbidden. However, that becomes
possible due to quantum effects. The computation results are shown in the
right part of Fig.\ref{Snake}: the first conductance peak is observed
below the threshold; above the threshold, the conductance exhibits an
oscillating behavior. The frequency corresponding to
the first conductance peak is close to the trapped frequency of the infinite
bent waveguide  (see Section \ref{4}) and can be thought as resonance frequency.

\section{Computation of trapped modes: brief review and new examples}\label{4}

The localized solutions  have been attracted a lot of attention
during last decades as examples of nonuniqueness to a scattering problem
which, at the same time, exhibit abnormal physical properties. Depending on the
area of application these solutions are either called trapped  modes
(microwave and/or acoustic waveguides, water waves), or bound states
(quantum wires and photonic crystals), or guided (Rayleigh-Bloch) waves
(diffraction gratings). The framework of a workshop publication doesn't suit
for detailed review of tens of papers devoted to the subject. The whole spectrum of
papers can be conventionally divided into three categories: proof of
existence or non-existence of trapped modes, asymptotic estimate of trapped
eigenfrequencies (energies) and numerical approaches. Not mention the first
category completely, we  point out the key  names and selected papers related to the
last two categories. The more comprehensive review in a part related to the quantum
applications can be found in the book \cite{LCM}.

Concerning waveguide problems, the asymptotic results can normally be obtained
if the geometry  is a perturbation of a straight strip. This may be a case
for small indentation \cite{BGRS}, laterally coupled waveguides through
small window(s) \cite{Popov}, slightly curved strips and tubes \cite{Exner},
etc. (see, e.g., \cite{KN1} for asymptotic results in gratings). Normally,
the asymptotic formulas match numerical computation for some range of
``small'' parameter; at the same time, numerics is often fails when
perturbation becomes very small.

As far as known to the author, the overwhelming majority of papers dedicated
to the numerical detection of trapped modes dealt with objects of a relatively simple
geometry. The typical methods were either the separation of variables in
sub-domains with subsequent matching of infinite series, or the reduction of
a problem to the integral equation (the explicit knowledge of Green's
function is required), or the application of a variational technique (the
sufficiently reach set of trial functions satisfying boundary conditions is
required). All these methods were successfully utilized and a wide variety
of trapped modes found by research groups represented by papers
\cite{LCM,Exner,Evans,McIver}.

However, all the mentioned methods are of limited universality and applicability
range.  The developed approach allows us to contribute into the
scope  providing successfully found new examples of trapped modes. Due to the
generality of the approach the list of examples can be very wide, below are
few of them. \footnotemark[5]\footnotetext[5]{The Workshop presentation
contains more ones, including examples related to the diffraction gratings.}

\begin{figure}[h]
\includegraphics[width=.45\textwidth]{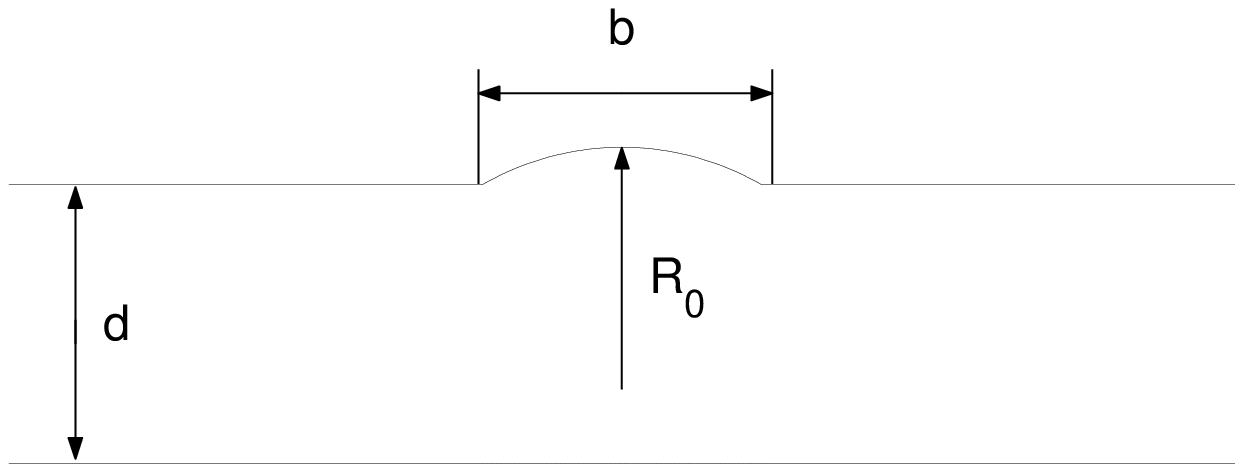}
\includegraphics[width=.5\textwidth]{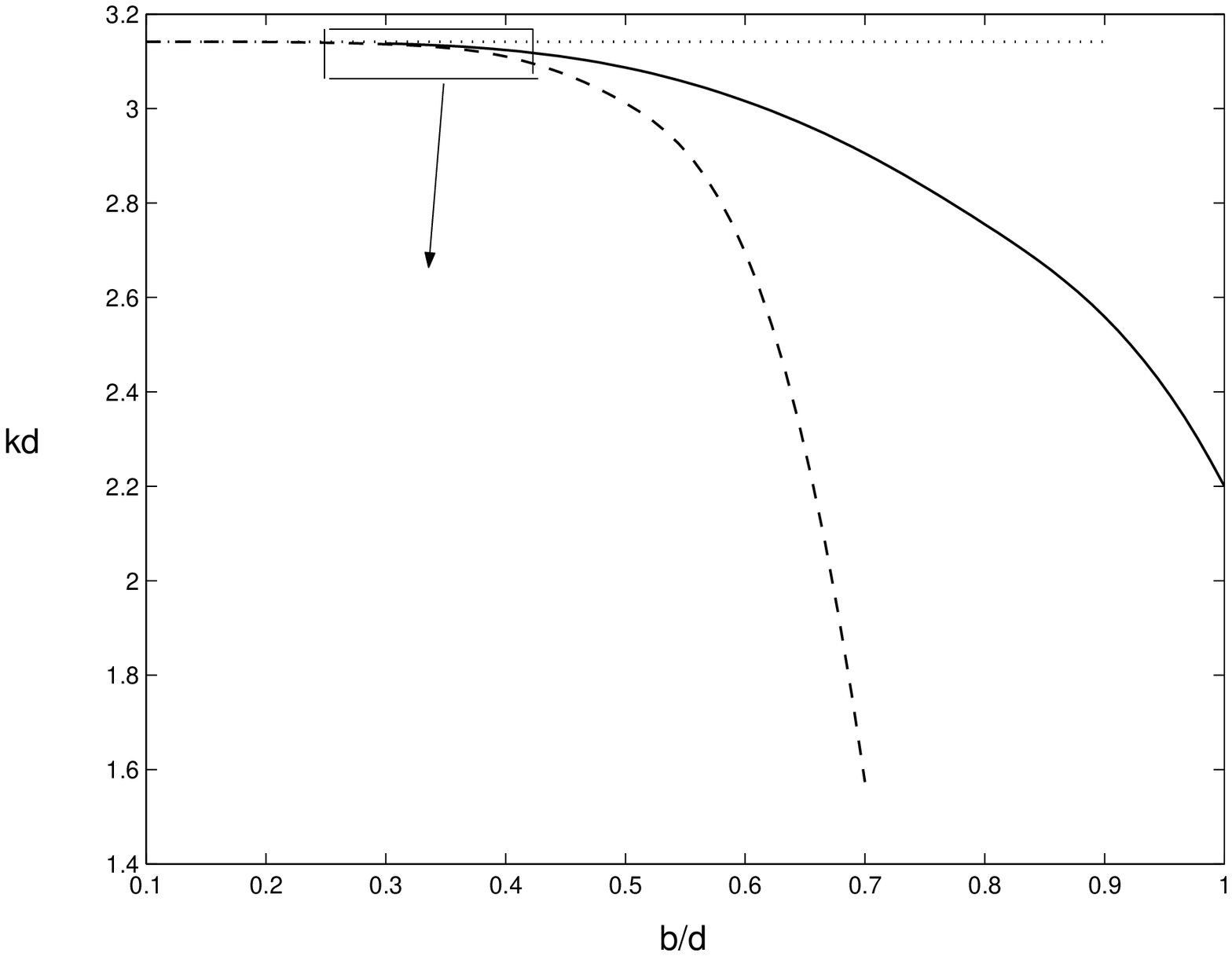}
\vskip -4cm
\hskip 9cm
\includegraphics[width=.16\textwidth]{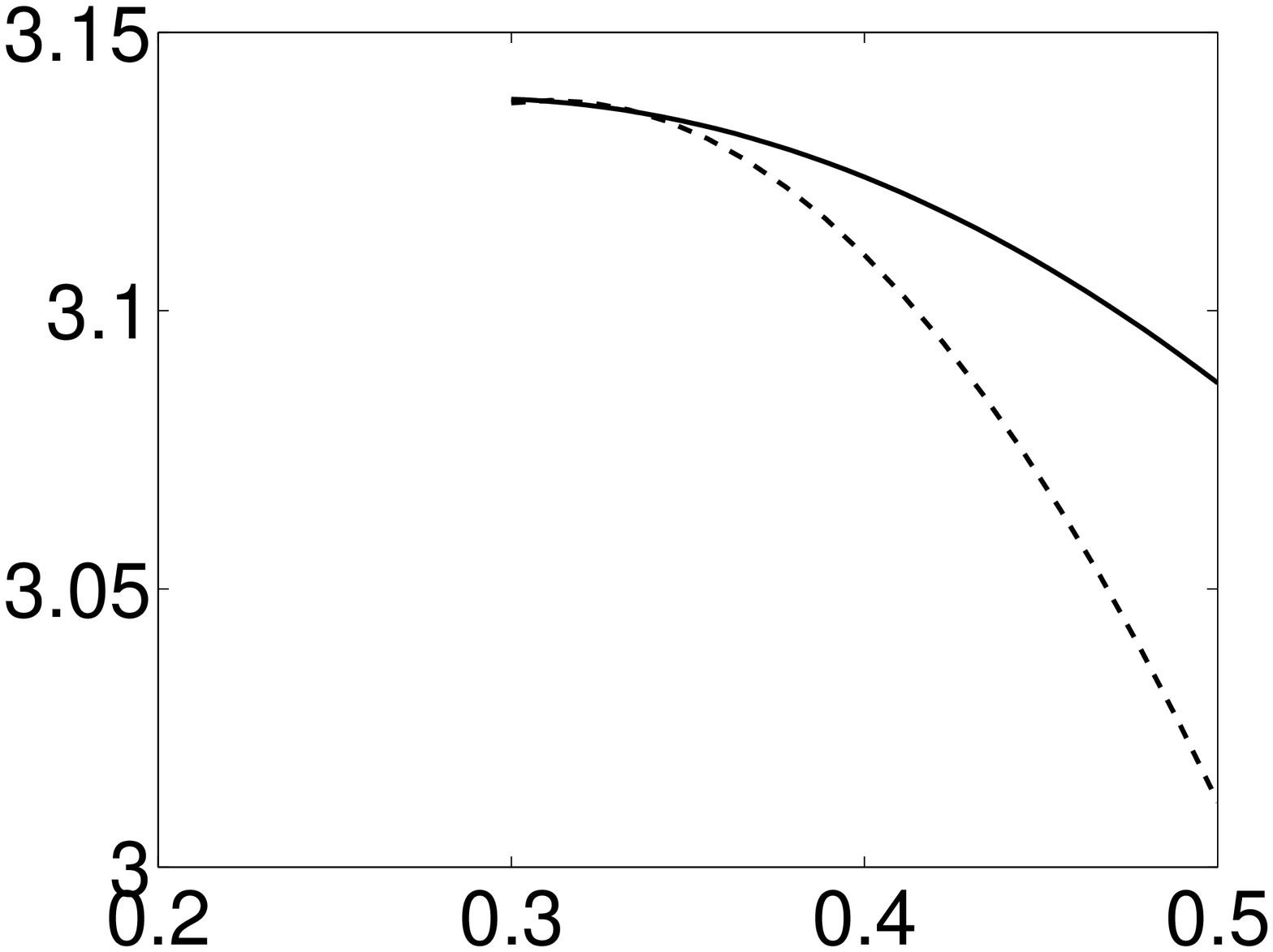}
\vskip 2cm
\caption{\small \textit{Left:} Waveguide with an one-side indentation.
\textit{Right:} The Dirichlet mode trapped eigenfrequencies $kd$ (solid line)
versus the length $b/d$ of the indentation (whilst $R_0/d = 1$)
in comparison with the asymptotic results of the paper
\cite{BGRS} (dashed line);
the inset shows the vicinity of the threshold in an expanded range.}
\label{One_side}
\end{figure}

The Dirichlet trapped modes eigenfrequencies for the straight waveguide with a local
indentation are shown in Fig.\ref{One_side}. Such model was
the subject of rigorous studies in \cite{BGRS} where existence of trapped
modes below the threshold were proved and asymptotic estimate of them found.
In Fig.\ref{One_side} the numerical results are given in comparison with the
asymptotic formulas  of \cite{BGRS}. It is seen that both results match each
other although the range of applicability of asymptotic formulas is
relatively narrow.

\begin{figure}[h]
\includegraphics[width=.45\textwidth]{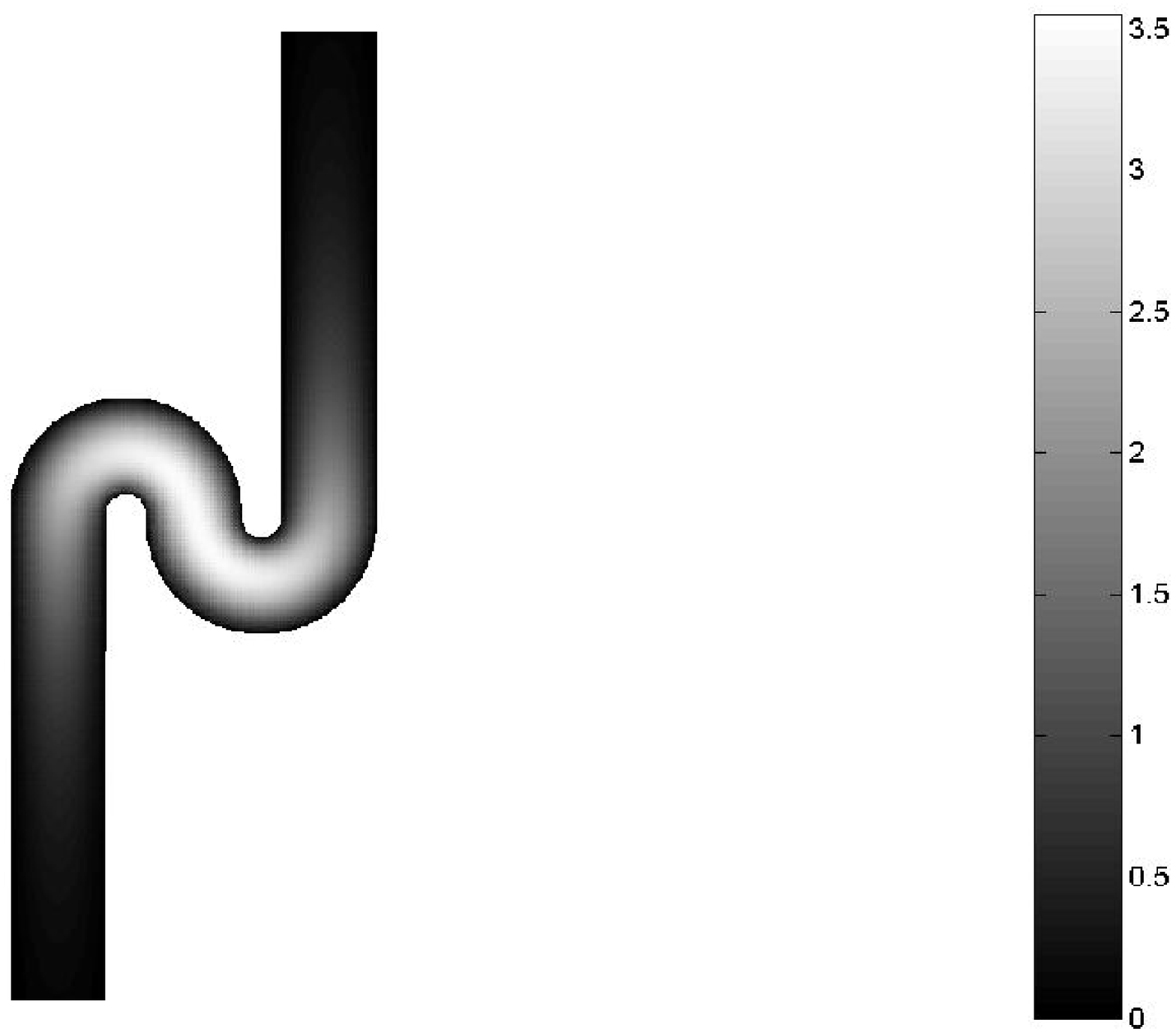}
\hspace*{-2cm}
\includegraphics[width=.55\textwidth]{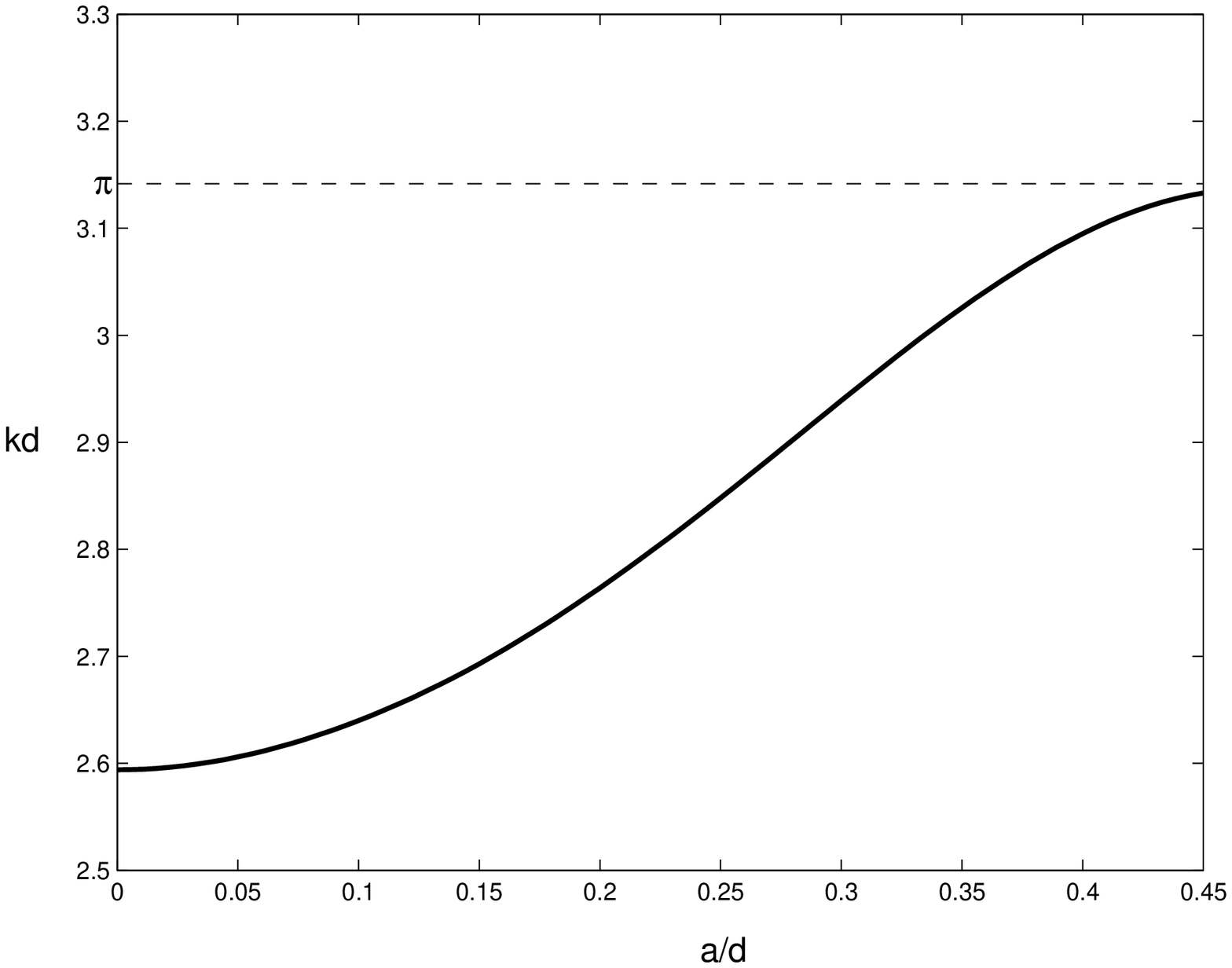}
\vskip -7cm
\hskip 7cm
\includegraphics[width=.2\textwidth]{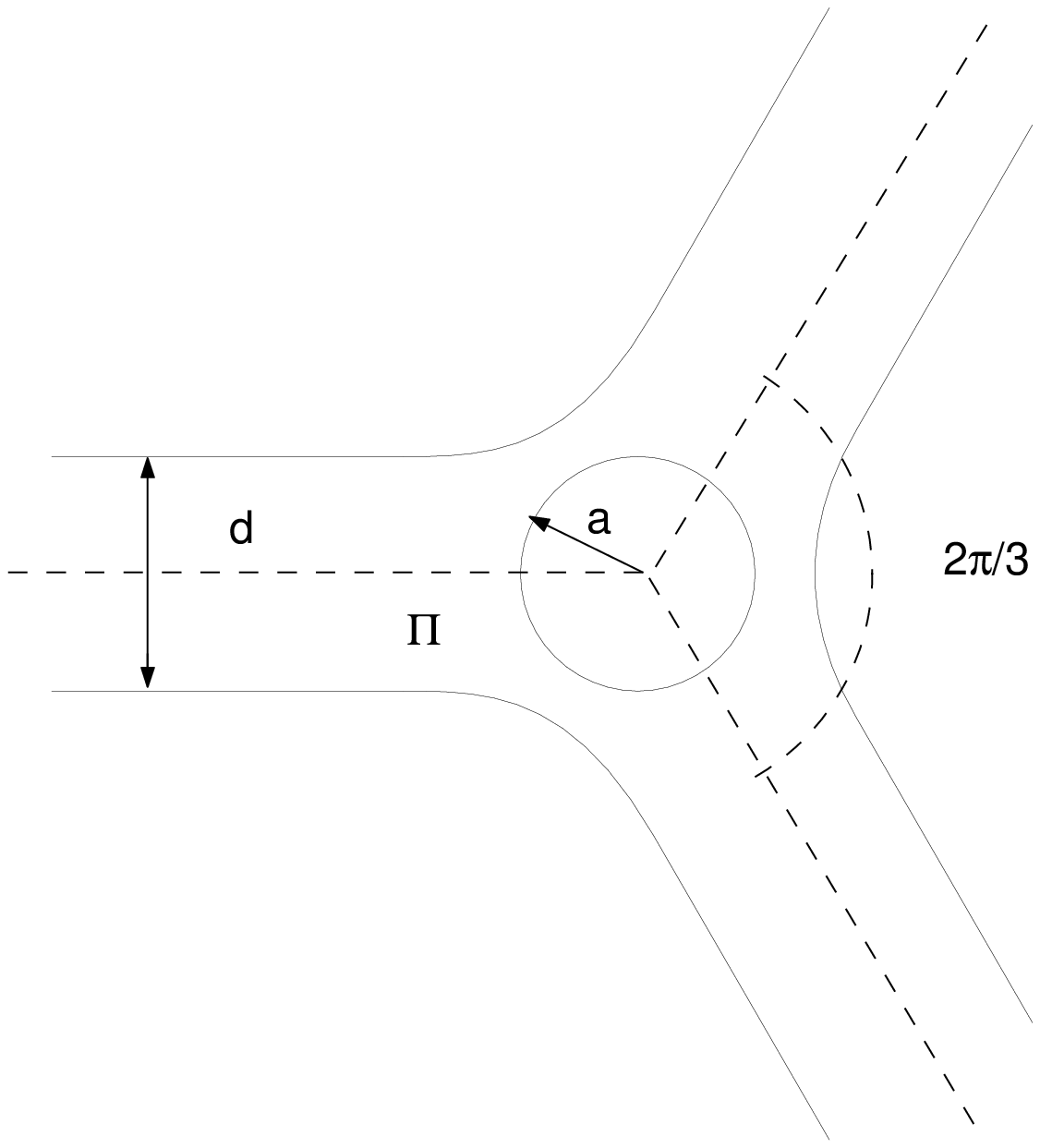}
\vskip 4cm
\caption{\small \textit{Left:} The intensity map of the trapped mode in a bent
waveguide; the bent part is the same as in Fig.\ref{Snake}, trapped mode
frequency is $kd\approx 3.091$ (which is slightly below the
sub-threshold conductance frequency for the bent waveguide with adjoint
leads, see  Section \ref{3.2}).
\textit{Right:} The dimensionless trapped eigenfrequencies $kd$
 (versus the size of disk hole) for the domain  shown in the inset.}
\label{Trapped}
\end{figure}

Examples in Fig.\ref{Trapped} concern the trapped modes for that no
numerical estimates were known. In the left part we show the
map of  $|\psi_{trapped}(x,y)|$ for a bent waveguide which bent structure is
the same as in the example of Fig.\ref{Snake} (but with no leads attached).
After replacing of the waveguide's straight parts by wider leads the
eigenfrequency (energy) of this trapped mode moves to a resonance in the
complex plane. This resonance works as a ``bridge'' providing the sub-threshold
conductance peak  in Fig.\ref{Snake}.

In the right part of Fig.\ref{Trapped} we provide an example of trapping
properties of the domain with relatively complicated geometry. The domain in
question consists of three smoothly
and symmetrically connected  channels (the symmetry and the number of channels
are not essential); some  disk is taken out around the symmetry
center. The Dirichlet boundary conditions are assumed at the outdoor domain
boundaries whilst the Neumann conditions are set at the interior circle
boundary. This domain with no hole ($a/d=0$) can be thought as a bent waveguide with an
``infinite'' indentation, thus the existence of the Dirichlet trapped mode
below the threshold
can be clearly predicted, and it is numerically found as $kd\approx 2.594$. The trapped
eigenfrequencies approaches the
threshold versus increasing the disk hole radius.

\section{Conclusion}

The general approach for numerical computation of scattering and trapping
properties of systems with finitely many channels to infinity quantum,
microwave and optics physics was discussed.
 The approach can be adopted for the numerical search of resonances
for the same type of scattering systems; this work is under development.

\end{document}